\documentclass{article}  
\usepackage{bigsky2009}
\usepackage{graphicx}
\frompage{000} \topage{000}                                              
\def\rightmark{Kinetic equilibration}
\def\leftmark{B. Zhang}
 
\title{Kinetic equilibration from a radiative transport}
\authors{
{Bin Zhang}\\[2.812mm]
{\normalsize
Arkansas State University, \\
PO Box 419, State University, AR 72467-0419, USA
}}
 
\abstract{Kinetic equilibration during the early stage
of a relativistic heavy ion collision is studied using a
radiative transport model. Thermalization is found to
dominate over expansion with medium regulated cross sections.
Pressure anisotropy shows an approximate $\alpha_s$ scaling
when radiative processes are included. It approaches an
asymptotic time evolution on a time scale of
1 to 2 fm/$c$. Energy density
is also found to approach an asymptotic time evolution that
decreases slower than the ideal hydro evolution.
These observations
indicate that viscosity is important during the early
longitudinal expansion phase of a relativistic heavy ion
collision.
}

 
\begin{document}
 
\maketitle
\setcounter{page}{1}

\section{Introduction}\label{intro}
One of the evidences for the discovery of the Quark-Gluon Plasma is
the collective behavior of produced particles that is consistent
with ideal hydro description of the evolution of the hot and dense
matter produced in relativistic heavy ion collisions. To go beyond
the ideal hydrodynamic description, viscous hydro has been 
developed \cite{Muronga:2001zk,Dusling:2007gi,Luzum:2008cw,Song:2007ux,Song:2008si,Song:2007fn}.
From a different perspective, microscopic transport models have
also been used to study the degree of thermalization 
\cite{Xu:2007aa,Zhang:2008zzk,Huovinen:2008te}. These studies show
that even small viscosity can affect the evolution and can be
important for the precise interpretation of experimental data.
In the following, we will describe kinetic equilibration of a
Quark-Gluon Plasma. In particular, we will discuss what happens
when medium regulated cross sections are used and when chemical
equilibration via particle number changing processes is included.
We will look at the central cell in central heavy ion collisions.
The kinetic equilibration will be characterized by the pressure
anisotropy. In addition, the energy density evolution will also be
studied. Finally, a summary will be given.

\section{Pressure anisotropy}

Local thermal equilibrium requires local pressure isotropy. For the
central cell in central heavy ion collisions, the degree of pressure
isotropization can be characterized by the longitudinal pressure
to transverse pressure ratio. Here the longitudinal direction is
the beam direction. Because of azimuthal symmetry, the two
transverse directions have the same pressure and the average will
be used as the transverse pressure. When the system is isotropic,
the longitudinal pressure is expected to be the same as the
transverse pressure, and the pressure anisotropy is expected to be
equal to one. We will first review findings from the 
\textsc{AMPT} model,
then we will introduce the new ingredients for the thermalization
study, i.e., medium regulated cross sections, and particle number
changing processes. Then we will look at how the two new
ingredients can help to maintain kinetic equilibrium by starting
with local thermal initial conditions. Furthermore, we will start
with zero initial pressure anisotropy as produced from an
inside-outside parton production process. This will give
information on how much equilibration a model can generate
with the inclusion of these two new ingredients.

Kinetic equilibration from the \textsc{AMPT} model was studied
in Ref.~\cite{Zhang:2008zzk}. The \textsc{AMPT}
model is a hybrid model that uses different models
for different stages of relativistic heavy ion 
collisions \cite{Zhang:1999bd,Lin:2000cx,Lin:2001yd,Lin:2004en}. 
It has two versions: the default model and the string melting
model.
The default model evolves minijets in \textsc{HIJING} 
\cite{Wang:1991ht} with \textsc{ZPC} \cite{Zhang:1997ej} and
reconnects them with their parent strings for fragmentation with
\textsc{PYTHIA} \cite{Sjostrand:1993yb}. 
The string melting model dissociates hadrons 
from \textsc{HIJING}
into their constituent quarks and anti-quarks. After the 
evolution of the quark-anti-quark plasma with \textsc{ZPC}, 
a coalescence model is
used to turn partons into hadrons. The hadron stage is handled
with the \textsc{ART} \cite{Li:1995pr} hadron cascade. 
The \textsc{AMPT} model has been used to study a variety of 
experimental observables \cite{zhang1,chen,zhang2,lin}.
The kinetic equilibration study 
yields a few interesting features. 
First, the string melting model has a
more rapid increase of the pressure anisotropy with proper time
as compared to the default model. 
The larger the cross section is,
the faster the increase of the pressure anisotropy. The pressure
anisotropy reaches a plateau. Later, as transverse expansion 
sets in, it increases and crosses one. It does not stay 
at one for any significant time period. This shows 
that only partial thermalization is achieved in these 
collisions.

The \textsc{AMPT} model has fixed cross sections for partons. In the
following, we will make use of medium dependent cross sections
and add particle number changing processes. The two gluons to
two gluons (2 to 2) cross section is taken to be the pQCD cross
section regulated by a medium generated Debye screening mass.
It is related to the screening mass by
\begin{equation}
\sigma_{22}=\frac{9\pi\alpha_s^2}{2\mu^2}.
\end{equation}
The 2 to 2 cross section $\sigma_{22}$ is proportional to the
strong coupling constant $\alpha_s$ squared and inversely
proportional to the screening mass $\mu$ squared. The screening
mass is related to $\alpha_s$, the cell volume $V$, and the
particle momenta via
\begin{equation}
\mu^2=\frac{8\pi\alpha_s}{V}\sum_i \frac{1}{p_i}.
\end{equation}
Notice that as the number of particles increases, even with the
same momentum distribution, $\mu^2$ increases. This leads to a
smaller 2 to 2 cross section. In other words, the collision rate
is regulated by screening, particles tend to interact with
close particles and can not reach very faraway particles.
This is a desirable feature and can avoid conceptual and technical
problems related to large cross sections in dense medium.

The simplest particle number changing processes will be included.
These are the 2 to 3 and 3 to 2 processes. The 2 to 3 cross section
is taken to be 50\% of the 2 to 2 cross section. This is roughly the
ratio obtained in a more elaborate model with pQCD 
matrix elements \cite{Xu:2004mz}.
The 3 to 2 rate is determined by detailed balance to ensure correct
chemical equilibration. The 3 to 2 reaction integral is related to
the 2 to 3 cross section by
\begin{equation}
I_{32}=192\pi^2\sigma_{23}.
\end{equation}
This relation assumes isotropic reactions. In the following, the
outgoing particle distribution will be taken as isotropic. This
is expected to catch the gross features of particle interactions
in the very dense central region while for the peripheral region
this may lead to deviations from results obtained from more
realistic forward type particle distributions.

We will start with local thermal initial conditions at an initial
proper time $\tau=0.5$ fm/$c$. In particular, two initial
temperatures ($T_0$) will be used, 0.5 GeV and 1 GeV. The
space-time rapidity will start uniformly 
within a range from -5 to +5.
The initial transverse coordinates will be within a disk of
radius of 5 fm. The space-time rapidity density is taken
to be $dN/d\eta_s=1000$. The low initial temperature case is
similar to what is expected at RHIC while the high temperature
case is a sketch of what may happen at the LHC. (Note that the
space-time rapidity density increases toward 2000 in the high
temperature case with its large initial energy density.) For
the early evolution, a time period from 0 to 2 fm/$c$ will
be used. This is early enough and no transverse expansion is
expected to affect the results. The evolution in the central
cell can be considered as having only longitudinal expansion.

\begin{figure}[!htbp]
\begin{minipage}{6.3cm}
\centering
\vspace{2mm}
\hspace{-5mm}
\includegraphics[scale=0.7]{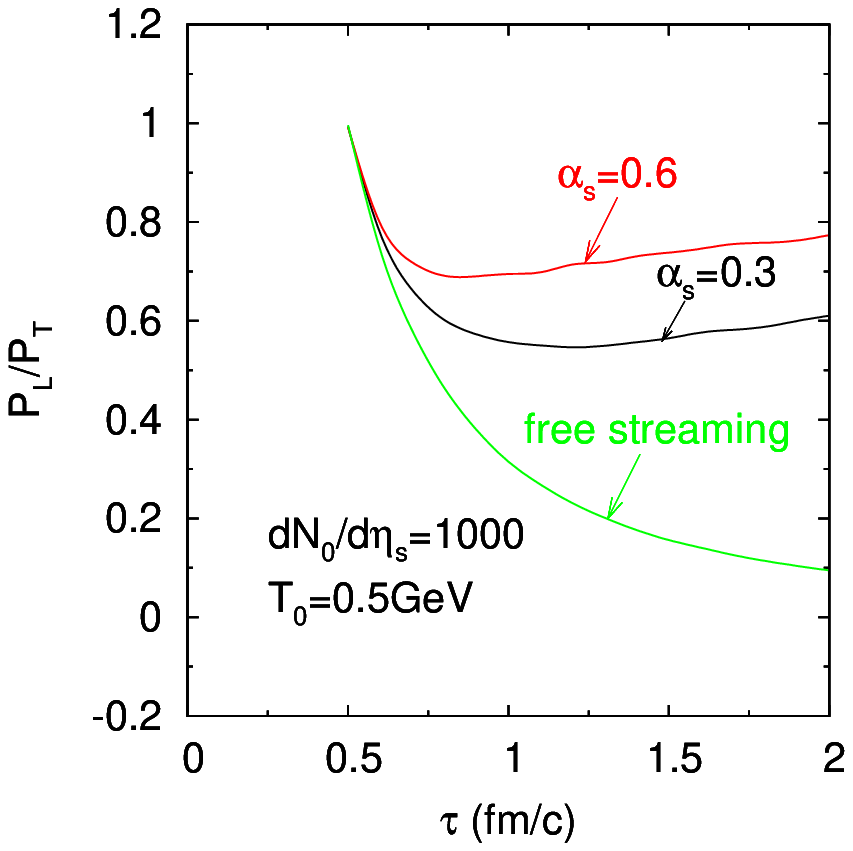}
\caption{\label{fig_plopt_iso1}
Pressure anisotropy evolution from local thermal initial 
conditions with different coupling constants.}
\end{minipage}
\begin{minipage}{6.3cm}
\centering
\hspace{-5mm}
\includegraphics[scale=0.7]{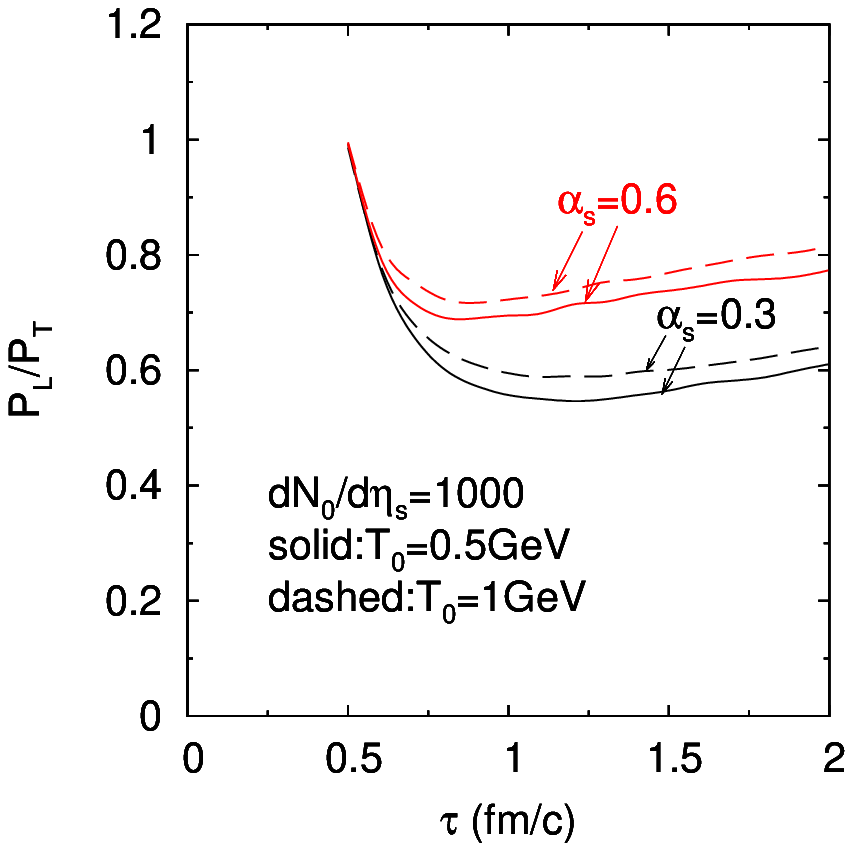}
\caption{\label{fig_plopt_iso2}
Like Fig.~\ref{fig_plopt_iso1} with additional
cases that have a higher initial temperature.}
\end{minipage}
\end{figure}

Fig.~\ref{fig_plopt_iso1} shows the proper time 
evolution of the pressure anisotropy
with the above set up. The free streaming curve can be used as a
guide of the effect of longitudinal expansion. It goes slightly
slower than $1/\tau^2$. As the interactions are turned on, the
decrease of pressure anisotropy slows down and after some time,
the pressure anisotropy reaches a minimum and then it increases
with time trying to return to isotropy again. 
As the coupling constant increases from 0.3 (the weak
coupling case) to 0.6 (the strong coupling case), the pressure
anisotropy goes closer to 1 and kinetic equilibration becomes
faster which can be seen from the fact that the minimum happens
earlier. A closer look at the evolution shows that if only the 2
to 2 process is included, there is an $\alpha_sT_0$ 
scaling \cite{Zhang:2008zzu}. If one evolution is solved, 
then the evolutions of systems with the same
$\alpha_sT_0$ are all solved. As 2 to 3 and 3 to 2 processes
are turned on, the evolution instead follows an approximate
$\alpha_s$ scaling (Fig.~\ref{fig_plopt_iso2}). 
For the same $\alpha_s$, the higher
the initial temperature, the closer the pressure anisotropy is
to one. However, as a result of the smaller cross sections
due to particle production,
the high initial temperature weak coupling
curve does not go high enough to match the low initial temperature
strong coupling case as in the case with only elastic collisions.

\begin{figure}[!htbp]
\begin{minipage}{6.3cm}
\centering
\vspace{2mm}
\hspace{-5mm}
\includegraphics[scale=0.7]{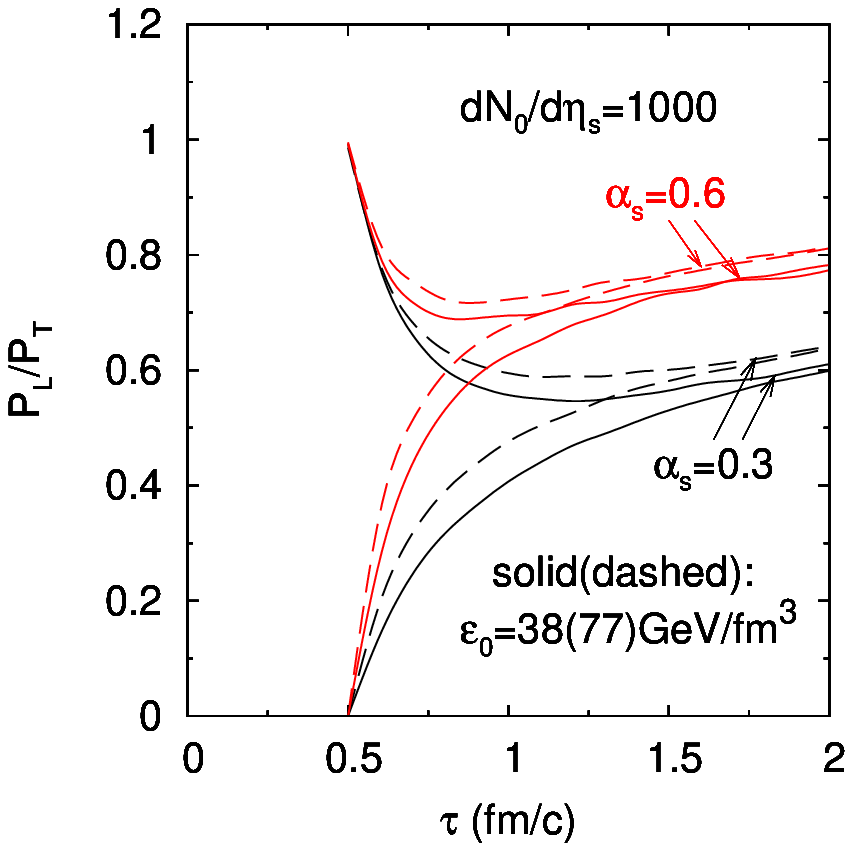}
\caption{\label{fig_plopt_iso_trans1}
Like Fig.~\ref{fig_plopt_iso2} with additional
cases that start with zero pressure anisotropy.} 
\end{minipage}
\begin{minipage}{6.3cm}
\centering
\hspace{-5mm}
\includegraphics[scale=0.7]{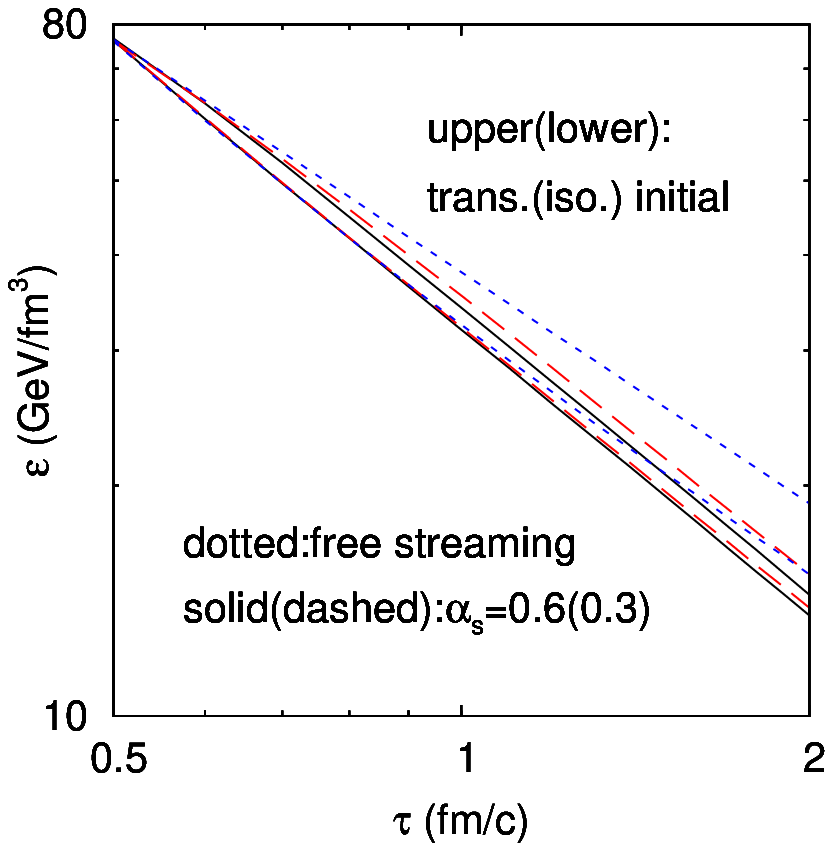}
\caption{\label{fig_eden1}
Energy density evolution on the log-log scale.
}
\end{minipage}
\end{figure}

Next we will look at what happens when there is no fast
thermalization. We will change the initial particle anisotropy.
At the initial time, the directions of particle momenta will be
changed so that they all move in the transverse plane. This is
consistent with the inside-outside particle production picture.
It can be compared with the case with an isotropic initial
particle distribution and the same initial energy density.
Fig.~\ref{fig_plopt_iso_trans1} shows that
the two cases with two different initial pressure anisotropies but
the same initial energy density move toward a common asymptotic
evolution. In other words, as the memory
of the initial condition fades away, the system evolves toward some
common evolution. The memory effect can be measured by the
difference between these two curves at any proper time relative
to the initial difference. 
A relaxation time can be used to describe how
fast the memory gets lost or how fast the isotropization is.
We notice that as the coupling constant increases,
the relaxation time decreases. Likewise, as the initial energy
density increases, the relaxation time decreases. Next we will
see if some common evolution also happens in the energy density
evolution.

\section{Energy density evolution}
For the evolution of azimuthally symmetric system of massless
particles, there are two independent thermodynamic quantities
for the central cell. In addition to the pressure anisotropy,
the energy density can be chosen as the other variable.

Fig.~\ref{fig_eden1} is a log-log plot of the energy 
density evolution. 
We will get a closer look at the case starting
with the high initial energy density. The low initial energy
density case shows roughly the same behavior. First, we notice
that the curves starting from the same initial energy density are
bounded by two curves, the upper dotted one is the free streaming
evolution from the transverse initial condition. The lower solid
one is the isotropic initial condition with strong coupling.
If particles free stream from the transverse initial
condition, they will stay within their cell when the system
expanses longitudinally. The total energy in the cell remains
the same while the volume increases due to longitudinal expansion.
The volume is proportional to the proper time. So the energy
density decreases as $1/\tau$. The evolution from an isotropic
initial condition with strong coupling constant can be fitted
by a power law. It gives $\epsilon\propto 1/\tau^{1.25}$.

The lower dotted curve is for free streaming from the 
isotropic initial condition.
It follows the isotropic initial plus strong coupling curve in
the beginning and deviates from it to approach the same power
law behavior for free streaming from the transverse initial
condition. Other curves also follow the early time power
law behavior which reflects the initial anisotropy. They also
develop asymptotic power law behavior that reflects interactions
at late times. The upper solid and upper dashed curves are from
the transverse initial condition with strong and weak interactions
respectively. They follow $1/\tau$ at early times and approach
$1/\tau^{1.25}$ at late times. The strong coupling case has
smaller energy density and the evolution is closer to the
strong coupling case from the isotropic initial condition.
The lower solid and dashed curves are for evolutions from
the isotropic initial condition with strong and weak
interactions respectively. Both follow $1/\tau^{1.25}$ approximately
while the strong coupling case has slightly smaller energy
density at late times. The time period that we studied is
about the time period when there is only longitudinal expansion
for our system. As from the pressure anisotropy study, this
is not much longer than the relaxation time. As a consequence,
the time evolution $1/\tau^{1.25}$ is slightly slower than
the ideal hydro $1/\tau^{1.33}$ energy density evolution.
Hence, viscous corrections are present and important during this
time period.

\section{Summary}

In this study, pressure anisotropy is shown to result from
the competition between longitudinal expansion and kinetic
equilibration. With medium regulated cross sections,
thermalization dominates over expansion at late times. There
is an approximate $\alpha_s$ scaling for the evolution of
pressure anisotropy when inelastic processes are included.
When evolving from the same initial energy density,
different initial pressure anisotropies follow the
same asymptotic
evolution when memory effect goes away. The relaxation
time is on the order of 1 to 2 fm/$c$. At the same time,
energy density also approaches the 
limiting power law evolution.
This evolution is slower than the 
ideal hydro evolution.
This indicates the importance of viscosity during the
early longitudinal expansion stage of evolution.
Further studies with more sophisticated initial conditions
and interactions will help to improve the estimate
of viscous effects and the impacts on experimental
observables.

\section*{Acknowledgments}
B.Z. thanks R. Bellwied, M. Guenther, U. Heinz, 
M. Lisa, U. Mosel, S. Pratt, I. Vitev 
for helpful discussions and
the Parallel Distributed Systems Facilities of the 
National Energy Research Scientific Computing Center
for providing computing resources.
This work was supported by the U.S. National Science Foundation
under grant No. PHY-0554930.

\vfill\eject

\begin{thebibliography}{99}

\bibitem{Muronga:2001zk}
  A.~Muronga,
  Phys.\ Rev.\ Lett.\  {\bf 88} (2002) 062302 
  [Erratum-ibid.\  {\bf 89} (2002) 159901]
  [arXiv:nucl-th/0104064].

\bibitem{Dusling:2007gi}
  K.~Dusling and D.~Teaney,
  Phys.\ Rev.\  C {\bf 77} (2008) 034905
  [arXiv:0710.5932 [nucl-th]].

\bibitem{Luzum:2008cw}
  M.~Luzum and P.~Romatschke,
  Phys.\ Rev.\  C {\bf 78} (2008) 034915
  [arXiv:0804.4015 [nucl-th]].

\bibitem{Song:2007ux}
  H.~Song and U.~W.~Heinz,
  Phys.\ Rev.\  C {\bf 77} (2008) 064901
  [arXiv:0712.3715 [nucl-th]].

\bibitem{Song:2008si}
  H.~Song and U.~W.~Heinz,
  Phys.\ Rev.\  C {\bf 78} (2008) 024902
  [arXiv:0805.1756 [nucl-th]].

\bibitem{Song:2007fn}
  H.~Song and U.~W.~Heinz,
  Phys.\ Lett.\  B {\bf 658} (2008) 279
  [arXiv:0709.0742 [nucl-th]].

\bibitem{Xu:2007aa}
  Z.~Xu and C.~Greiner,
  Phys.\ Rev.\  C {\bf 76} (2007) 024911
  [arXiv:hep-ph/0703233].

\bibitem{Zhang:2008zzk}
  B.~Zhang, L.~W.~Chen and C.~M.~Ko,
  J.\ Phys.\ G {\bf 35} (2008) 065103
  [arXiv:0705.3968 [nucl-th]].

\bibitem{Huovinen:2008te}
  P.~Huovinen and D.~Molnar,
  arXiv:0808.0953 [nucl-th].

\bibitem{Zhang:1999bd} B.~Zhang, C.~M.~Ko, B.~A.~Li and Z.~W.~Lin,
Phys.\ Rev.\ C \textbf{61} (2000) 067901 [arXiv:nucl-th/9907017].

\bibitem{Lin:2000cx} Z.~W.~Lin, S.~Pal, C.~M.~Ko, B.~A.~Li and B.~Zhang,
Phys.\ Rev.\ C \textbf{64} (2001) 011902(R) [arXiv:nucl-th/0011059].

\bibitem{Lin:2001yd} Z.~W.~Lin, S.~Pal, C.~M.~Ko, B.~A.~Li and B.~Zhang,
Nucl.\ Phys.\ A \textbf{698} (2002) 375 [arXiv:nucl-th/0105044].

\bibitem{Lin:2004en} Z.~W.~Lin, C.~M.~Ko, B.~A.~Li, B.~Zhang and S.~Pal,
Phys.\ Rev.\ C \textbf{72} (2005) 064901 [arXiv:nucl-th/0411110].

\bibitem{Wang:1991ht} X.~N.~Wang and M.~Gyulassy,
Phys.\ Rev.\ D \textbf{44} (1991) 3501.

\bibitem{Zhang:1997ej} B.~Zhang,
Comput.\ Phys.\ Commun.\ \textbf{109} (1998) 193 [arXiv:nucl-th/9709009].

\bibitem{Sjostrand:1993yb} T.~Sjostrand,
Comput.\ Phys.\ Commun.\ \textbf{82} (1994) 74.

\bibitem{Li:1995pr} B.~A.~Li and C.~M.~Ko,
Phys.\ Rev.\ C \textbf{52} (1995) 2037 [arXiv:nucl-th/9505016].

\bibitem{zhang1}B. Zhang {\it et al.}, Phys. Rev. C {\bf 62} (2000)
054905; {\it ibid.} {\bf 65} (2002) 054909.

\bibitem{chen}L. W. Chen and C. M. Ko, Phys. Rev. C {\bf 73} (2006) 044903.

\bibitem{zhang2}B. Zhang, L. W. Chen and C. M. Ko, Phys. Rev. C {\bf 72}
(2005) 024906.

\bibitem{lin}Z. W. Lin, C. M. Ko and S. Pal, Phys. Rev.
Lett. {\bf 89} (2002) 152301.

\bibitem{Xu:2004mz}
  Z.~Xu and C.~Greiner,
  Phys.\ Rev.\  C {\bf 71} (2005) 064901
  [arXiv:hep-ph/0406278].

\bibitem{Zhang:2008zzu}
  B.~Zhang,
  arXiv:0809.0446 [nucl-th].


\end{thebibliography}
\end{document}